# Topologically integrated photonic biosensor circuits

Ze-Lin Kong[1, *], Yang Liu[1, *], and Jian-Hua Jiang[2, 3, 4, 1, †]

[1]School of Physical Science and Technology & Collaborative Innovation Center of Suzhou Nano Science and Technology, Soochow University, Suzhou 215006, China

[2]School of Biomedical Engineering, Division of Life Sciences and Medicine, University of Science and Technology of China, Hefei 230026, China

[3]Suzhou Institute for Advanced Research, University of Science and Technology of China, Suzhou 215123, China

[4]Department of Modern Physics, School of Physical Sciences, University of Science and Technology of China, Hefei 230026, China

*These authors contributed equally to this work.

†Correspondence to: jhjiang3@ustc.edu.cn (JHJ)

**Abstract**

Integrated nanophotonic biosensors offer a promising route toward future biomedical detection applications that may enable inexpensive, portable, and sensitive diagnosis of diseases with a small amount of biological samples for convenient early-stage screening of fatal diseases. However, the current photonic biosensor designs are not suitable for highly integrated and multiplexing device architectures that can achieve the detection of complex combinations of many biomarkers. Here, we propose a topological scheme for the integration of miniature biosensors in photonic crystal chips that can meet the above requirement. Using photonic topological edge states as robust one-dimensional waveguides that connect many photonic biosensors, we propose here the topologically integrated photonic biosensor circuits. We demonstrate that the performance of the topologically integrated photonic biosensors is much more robust against disorders than that of the photonic biosensors connected by the normal photonic waveguides, due to the robust transport of photons along the edge channel. Since disorders arising from the fabrication imperfection and the random distribution of the biomarkers are inevitable in genuine devices, resilience against disorders is a necessity for on-chip integration of biosensors. The topological scheme proposed here thus opens a promising path toward reliable integration of photonic biosensors for next-generation biomedical applications.

**Introduction**

The growing needs of affordable medical care demand for advanced technologies that can provide portable, convenient, and inexpensive medical testing and diagnosis at home to screen potential diseases in the pre-hospital stage. Photonic biosensors are potentially a solution to satisfying such social and technology needs. With machine learning based on big data and the many biomarkers in the photonic biosensors as well as efficient measurements, it is possible to screen fatal diseases at the early stage without relying on time- and resource-consuming hospital processes. Moreover, due to the miniature nature of the photonic biosensors, the biological specimens needed is very small. However, such a blueprint of advanced biomedical technology relies on massive integration of reliable photonic biosensors in a single chip, which is still very challenging now.

Conventional photonic biosensor schemes are based on the refractive index sensing caused by the disease markers at isolated photonic cavities which often suffer low level of integration and performance. In the past decade, much research efforts have been devoted to integrated photonic biosensors. Along this direction, various structures such as microring resonators [1-4], Mach-Zehnder interferometers [5], surface plasmons resonators [6], and photonic crystal (PhC) cavities are designed to realize integrated photonic biosensors [7-13]. These schemes are, however, still vulnerable to inevitable fabrication imperfections.

Here, we propose a novel scheme to connect the photonic cavities together to form an integrated, multiplexing photonic biosensor circuits. The uniqueness of our scheme is to couple the biosensors with topological edge states in photonic crystals (PhCs). The essential mechanism underlying this scheme is the robust transmission properties of the coupled edge-cavity system. In this regard, we find that the transmission spectrum of the normal coupled waveguide-cavity system is vulnerable to defects and disorders. In contrast, the transmission property of the coupled edge-cavity system is sensitive only to the perturbations at the cavity (i.e., the biosensor). Therefore, the performance of such photonic biosensor is stable against the fabrication errors away from the biosensor. Moreover, the unavoidable distribution of the biomarkers away from the biosensor do not affect the transmission spectrum of the edge-cavity coupled system. We further demonstrate that by coupling many biosensors to the same edge channel, an integrated photonic biosensor circuit with robust, multiplexing biosensing can be formed. The scheme proposed here holds the promise for high-performance on-chip integration of photonic biosensor circuits that may meet the needs for high-throughput biosensing in future biomedical technology.

**Main results**

The basic element of the proposed scheme is the coupled edge-cavity system where the photonic topological edge channel is coupled with a photonic crystal defect cavity to form a waveguide-sensor system. Many such elements can be connected in series or in parallel to form the integrated photonic circuits as the highly integrated photonic biosensors chip that can serve as powerful diseases diagnosis. We will demonstrate that such a scheme has a number of advantages: First, the basic element has high sensitivity and reliable performance due to topological protection of the edge channel. Second, the resilience of such topological photonic biosensors enables robust integration of them via serial couplings. Remarkably, we find that through the above design, the crosstalk and interference between these serially coupled photonic biosensors are significantly reduced. These advantages inspire us to propose the concept of topologically integrated photonic biosensors where the above topological photonic biosensors can be robustly integrated in a single chip via both serial and parallel connections that enable a massive network of photonic biosensors for next-generation biomedical diagnosis technology [14,15] (see Fig. 1).

We illustrate the advantages of the topological photonic biosensors by benchmarking with the conventional photonic crystal biosensors where photonic crystal waveguides (PCWs) are used in place of the photonic topological edge channels. In the conventional scheme, the sensing function is realized by many isolated photonic cavities where the biorecognition mechanisms such as the antibody-antigen binding and DNA aptamer-protein binding [16,17] are used to trap the disease markers. These disease markers are then detected through the resonance frequency shift of the photonic cavities via the index sensing mechanism. Thus, each photonic cavity realizes a photonic biosensor. Such schemes often suffer from low integration density and other issues, although they may outperform the conventional biosensing technology. To make high-performance photonic biosensors, we target at highly integrated optical circuits of miniature photonic biosensors with multiplexing functions.

In this work, we provide such a design based on the coupled edge-cavity systems. Our design scheme has the following advantages: First, the coupled edge-cavity system has quite robust transmission properties. That is, when disorders (such as those induced by fabrication errors) are introduced into our system, the transmission spectrum remains almost the same, except when the disorders are around the cavity which is designed for sharp sensing of refractive index changes induced by disease markers. This robust transmission property due to the photonic topological edge states is the key feature for our scheme. For instance, the robust transmission sets the ground for us to couple many photonic cavities with the same edge channel while keeping the independence and stableness of the performance of each photonic biosensor. As shown in Fig. 1(a), we can couple four photonic cavities with one edge channel where each photonic cavity has a particular frequency. In the transmission spectrum, the resonance frequency of a photonic cavity is identified as a dip in the transmission with vanishing transmittance [Fig. 1(b)]. Moreover, as shown in Fig. 1(c), there is negligible interplay and crosstalk between the transmission of each individual photonic cavity when they are coupled with the same edge channel. For instance, we demonstrate in Fig. 1(b) that although sharing the same topological edge channel with three other photonic cavities, the transmission spectrum of a photonic cavity is nearly the same as when there are no other photonic cavities.

In our design, we consider slab photonic crystals which are feasible in the current fabrication technology and can provide a large photonic band gap to allow sufficient integration of photonic cavities with different resonance frequencies. Moreover, the photonic cavities based on slab photonic crystals can achieve high quality factors as shown by the previous study [18]. High quality factors of the photonic cavities could enhance the sensitivity of the photonic biosensors. When multiple photonic biosensors are coupled to the topological photonic edge channel, the quantitative characterization of the biomarkers at many photonic biosensors can be extracted simultaneously by a single transmission measurement. Our scheme thus realizes multiplexing of photonic biosensors using the frequency domain. Furthermore, many such photonic crystal chips can be integrated in the same biosensing device as illustrated in Fig. 1(d) which further enhances the level of integration and multiplexing. With reliable micro-fabrications, this scheme can lead to massively integrated photonic circuits. Since such an integration scheme relies fundamentally on the topological edge channels, we term this scheme as topologically integrated photonic biosensor circuits. In the following sections, we will present systematically the design principles and details, the integration scheme, and the performances of the topologically integrated photonic biosensors.

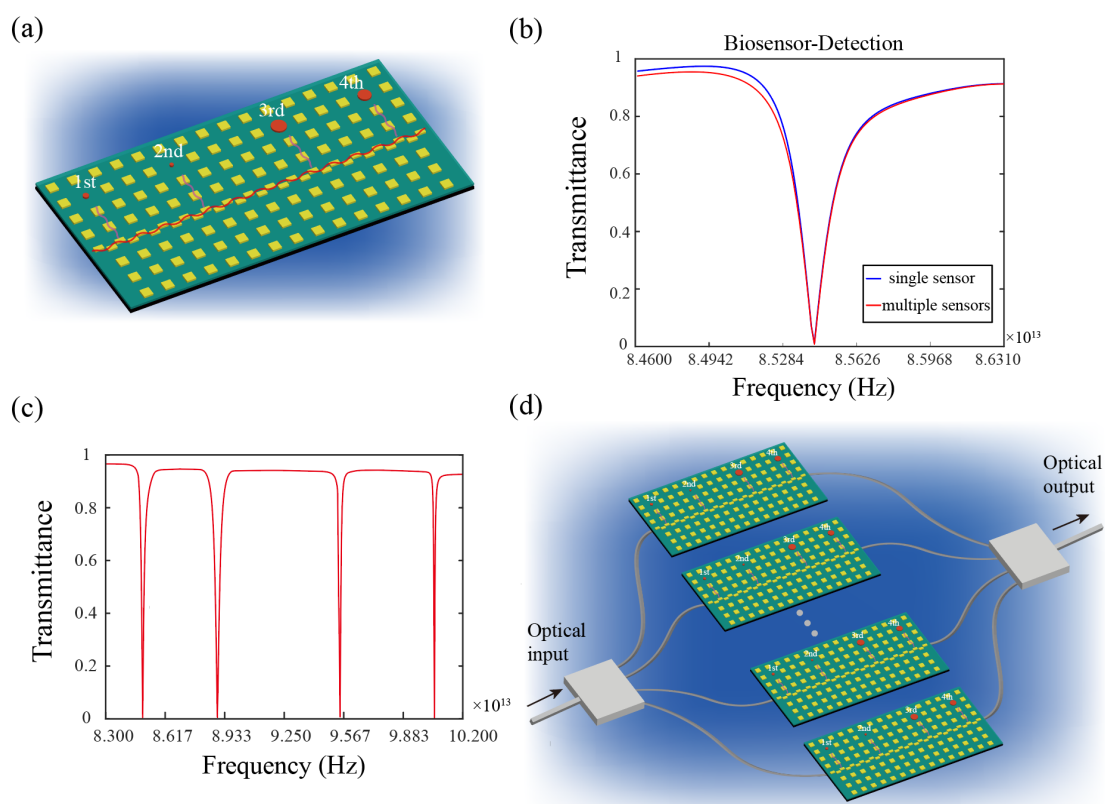

**Figure 1.** (a) Schematics of topologically-integrated photonic biosensors based on dielectric PhCs. Yellow squares denote the dielectric square rods that form the PhCs. Red squares denote the shrunk or expanded dielectric square rods that create defect-like photonic cavities which serve as photonic biosensors via biomarker binding mechanisms and refractive index sensing. The wavy lines indicate

the topological edge channel and the couplings between the photonic cavities with the edge channel. (b) Illustration of the topology-enabled robustness of the signature transmission spectrum of the photonic biosensor chip where two cases are compared: In one case, there is only a single biosensor, while in the other case, there are multiple biosensors coupled serially to the same edge channel. (c) Illustration of a typical transmission spectrum of the topologically-integrated photonic biosensor chip where multiple photonic biosensors coupled serially to the same edge channel. (d) Schematic of photonic biosensor circuits where many photonic biosensors are integrated by topological edge channels serially and parallelly. This topologically-integrated photonic biosensor circuits can enable massive measurements of disease markers via a single-shot transmission measurement.

**Photonic crystal design**

PhCs comprised of periodic structures of dielectric materials provide remarkable tools for the manipulation of light. The emergence of photonic bandgaps in PhCs [19-32] give access to versatile design of photonic cavities and waveguides. Therefore, PhCs offer an appealing motherboard for the design of photonic biosensors. We shall illustrate our scheme based on dielectric PhCs. However, the proposed scheme also applies to other photonic periodic structures.

To realize the topologically integrated photonic biosensor circuits, we first design two-dimensional (2D) PhCs to form the photonic topological edge channels and the miniature photonic cavities. Although the PhCs will eventually be in the slab geometry for realistic fabrications, here we consider them ideally as 2D PhCs to unveil our design principles in a simple setup. The photonic topological edge states are designed via the inspiration from the recent studies on topological Wannier cycles [33, 34]. We consider the interfaces between two photonic crystals with different Wannier centers. If an interface has the glide symmetry, it can induce efficiently artificial gauge flux into the photonic system and thus leads to photonic topological edge states without breaking the time-reversal symmetry [33]. Topological Wannier cycles can provide an efficient way toward photonic topological edge states in 2D dielectric PhCs at optical and infrared frequencies where time-reversal symmetry breaking is hard to achieve.

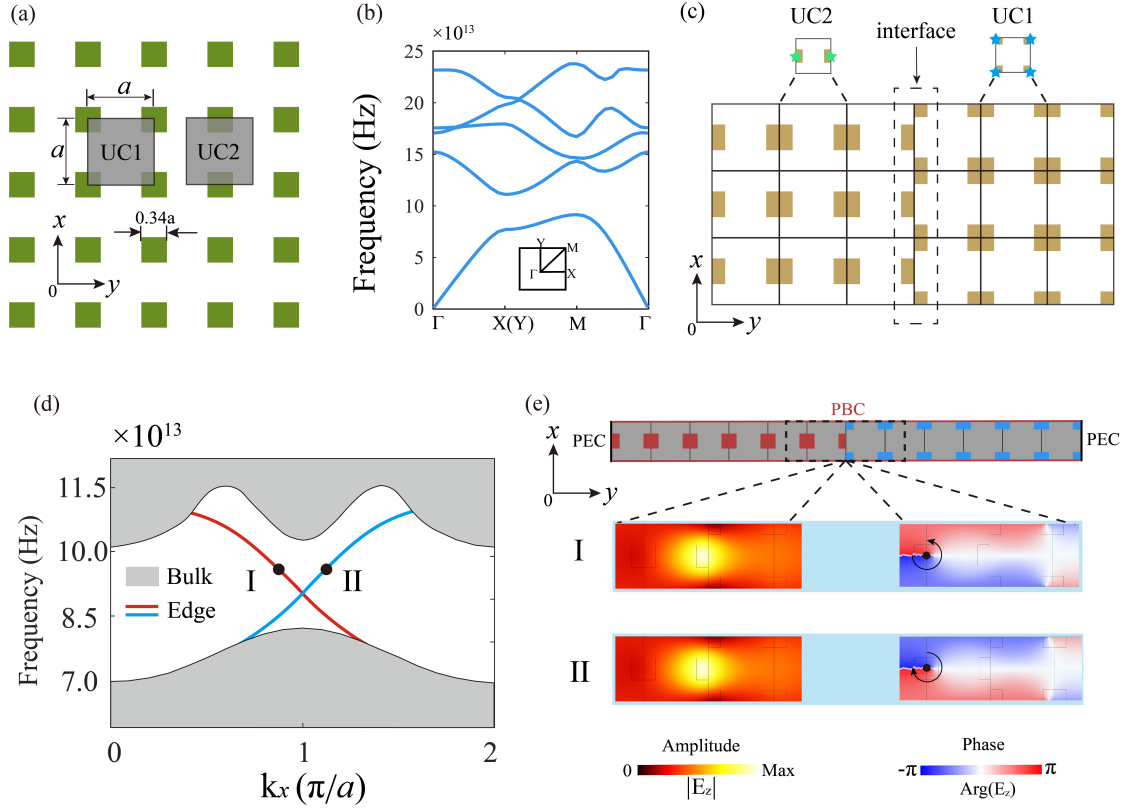

**Fig. 2.** (a) Schematics of the dielectric square lattice PhC. The dielectric rods have a side length $d = 0.34a$ and a relative permittivity $\varepsilon = 11.56$. Two choices of unit-cells are depicted as UC1 and UC2, respectively. (b) Photonic bands along the high symmetry lines of the first Brillouin zone for the PhC in (a). Inset illustrates the first Brillouin zone. (c) In the top-right diagram the Wannier centers for UC1 are represented by the blue stars, while the top-left diagram illustrates the Wannier centers for UC2, as represented by the green stars. Here only the first photonic band is concerned. The main figure depicts a supercell which is periodic in the *y* direction and exhibit a glide reflection symmetry around the interface in the middle. (d) The projected photonic band structure along the *x* direction for the supercell structure shown in (e). Gray regions are the bulk bands, and the red (blue) lines give the edge states dispersion. (e) Schematics of the supercell structure composed of two square lattice PhCs as well as the distributions of the amplitude and phase of the electric field for the edge states highlighted by black circles in (d). Results here are calculated via the eigenstates solutions to the Maxwell equations using finite-element methods.

With such a scheme, the photonic topological edge states can be realized at the interface between two square lattice PhCs, as shown in Fig. 2. We consider the 2D PhCs made of silicon square rods with a side length $d = 0.34a$ and a dielectric constant

$\varepsilon = 11.56$, which is arranged in water to form square lattices. The lattice constant of the PhCs is $a = 1\mu m$. The schematic of the square lattice PhCs is shown in Fig. 2(a) where the olive squares represent the dielectric square rods. Two choices of unit-cells (labeled as UC1 and UC2, respectively) are considered [Fig. 2(a)], leading to two types of PhCs. They have the same photonic band structure [Fig. 2(b)] but different Wannier centers for the first photonic band [Fig. 2(c)]. Therefore, the two configurations, UC1 and UC2, lead to photonic bands of distinct topological properties. The underlying physics is similar to the Wannier centers of the Su-Schrieffer-Heeger model in the trivial and topological phases: The Wannier centers in these two phases are shifted by half of the lattice constant. Moreover, as illustrated in Fig. 2(c), the interface between the two PhCs has a glide symmetry: By taking a mirror reflection with respect to the interface and a half-lattice-constant translation along the interface boundary, the geometry of the supercell comes back to itself. This glide symmetry of the supercell is essential for the emergence of the gapless interface states in the photonic band gap, as revealed below.

As discovered in Ref. [33], such an interface can be regarded as a source of the artificial gauge flux on the photonic states which drives the system to achieve cyclic pumping between the photonic states of even and odd parities. Such cyclic pumping eventually generates a pair of topological edge states traversing the whole photonic band gap, as shown in Fig. 2(d). One can see that the two branches of edge states are related by the time-reversal operation. In fact, we find that the two topological edge states I and II [labeled by the black dots in Fig. 2(d)] support photonic orbital angular momenta of the opposite values [as indicated by the identical field amplitude and the opposite phase windings in Fig. 2(e)]. This feature, which is similar to the spin-momentum locking in the edge states of the quantum spin Hall insulators, clearly illustrates the relationship between the edge states I and II as the time-reversal partners of each other.

As the topological edge states traverse the entire photonic band gap, they provide a one-dimensional (1D) channel that covers a wide frequency range---a channel that can serve as a robust information bus line that carries the information at each photonic biosensor to the output terminal. The inclusion of such a 1D channel is a key feature of our design. The typical behavior of the system when a photonic biosensor is coupled to the edge channel is a dip in the transmission at the resonance frequency of the biosensor, as shown in Fig. 1 as well as in previous studies [7-13]. Tailoring the distance between the cavity and the interface can control the width of the transmission dip. As shown in Figs. 3(a-c), when the distance of cavity and interface is equal to $4a$, the calculation results show that the transmission spectrum is sharper, giving a transmission curve that is promising for the integrated photonic biosensors. In ideal 2D PhCs, the defect-like cavity states can have very long lifetime, if it is not coupled with other states. Therefore, the sharpness of the transmission dip can be further increased by enlarging the distance between the cavity and the interface, i.e., reducing the coupling between the cavity mode and the interface states. In genuine 2D slab PhCs, other mechanisms can limit the quality factor of the defect-like cavities. Thus, there will be an optimal value of the cavity-interface distance that gives an optimal value of the cavity-edge coupling that are best for the sensitivity of the photonic biosensor as well as the depth of the transmission dip. In the following, we shall use the cavity-interface distance $4a$ as the basic setting for the quantitative discussions of the performance of the designed topological photonic biosensor chip.

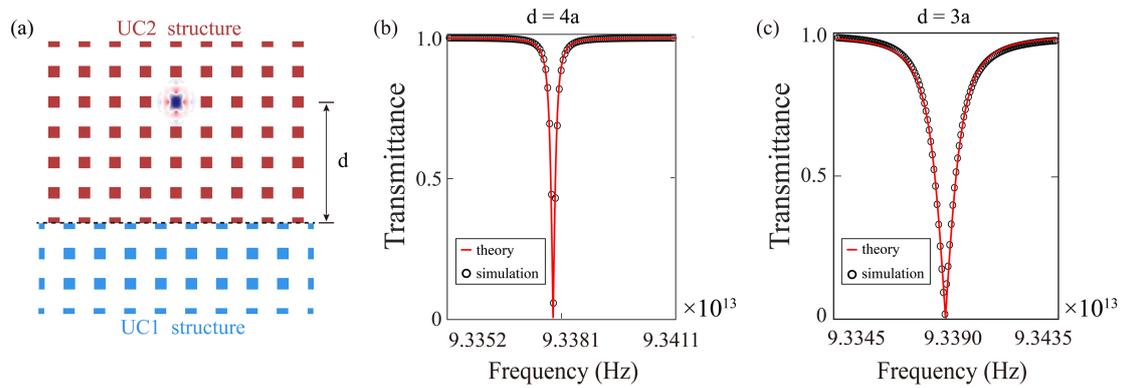

**Figure 3.** (a) Schematic of the coupling between the photonic cavity and the edge states at the interface when their distance is $d$. (b) The transmission spectrum for the structure with distance $d = 4a$. (c) The transmission spectrum for the structure with distance $d = 3a$. Here, the incident wave is coming from the left, while the transmitted wave is on the right side.

In this work, two approaches are employed to calculate the transmission spectrum.

In one approach (denoted as "simulation"), the transmission spectrum can be calculated by directly calculating the energy flux of photons at the left and right boundaries of the simulated photonic structure. The left boundary gives the incoming and reflected photon energy flux, while the right boundary gives the outcoming photon energy flux. The transmission is then obtained as the ratio between the outcoming and the incoming photon energy fluxes. In another approach (denoted as "theory"), one can match the photonic wave amplitudes at different positions with the plane wave ansatz of the incoming, outcoming and reflected waves. By taking several positions, one can have enough conditions to solve the problem and determine the amplitude of the transmitted photonic waves from which the transmission spectrum can be obtained. In all cases, the above two approaches give the same results, as illustrated in Figs. 3(b) and 3(c). In the following sections, we shall illustrate only the transmission spectrum from the second approach.

To illustrate the appealing properties of our topological photonic biosensors [Fig. 4(a)], we consider the effects of disorder on the performance of the photonic biosensors. For this purpose, we choose a conventional photonic biosensor based on the coupling between a defect-like cavity and a PhC waveguide (PCW) based on a vacant channel in the PhC [Fig. 4(c)]. For the fairness of comparison, we set all parameters of the PhC and the defect-like cavity the same as in the topological case. The distance between the cavity and the PCW is also chosen as $4a$. In addition, the PCW states have a dispersion spreading the whole photonic band gap [Fig. 4(d)]. To reveal the effects of the disorder, we calculate the transmission spectra with and without the disorder for both the normal and the topological biosensors. For each calculation, the disorder configurations are the same for both biosensors. Specifically, the disorders are simulated by randomly putting small dielectric rods near the center of the waveguide channel for the normal case and the interface between the UC1 and UC2 PhCs for the topological case.

Remarkably, we find that the transmission spectrum of the topological biosensor is much more robust against disorders than the normal biosensor [Figs. 4(a) and (e)]. This result indicates that the performance of the topological biosensor is more stable than the normal biosensor, when considering the fabrication errors and randomly distributed biomaterials under genuine working conditions. We remark that such stableness should not be mixed with the sensitivity of the biosensor which comes from the refractive index sensing near the defect-like cavity.

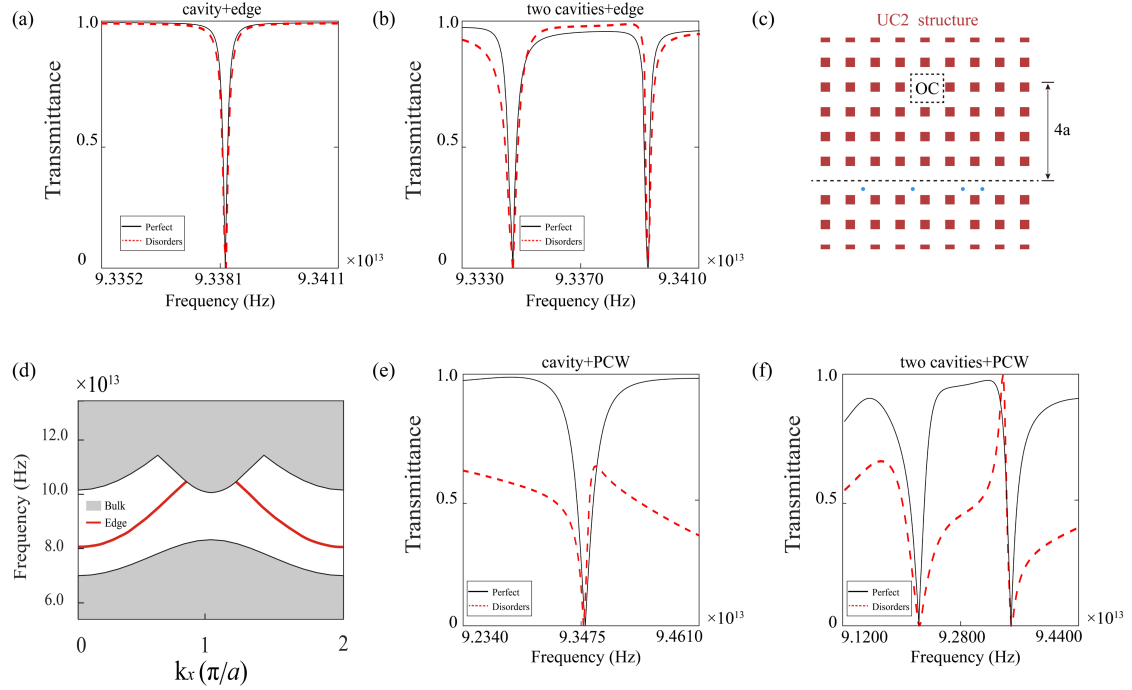

**Fig. 4.** (a) The transmission spectra of photonic edge state structure. The solid black line denotes an ideal boundary, while the red dashed line indicates the introduction of defects at the boundary. (b) The transmission spectra of edge states in a serial-coupled optical cavity system. (c) Photonic cavity waveguide structure schematic diagram. The entire structure is designated as UC2. By eliminating a row of dielectric rods, the boundary is established as an empty waveguide, with a distance of 4a between the cavity and the waveguide. (d) The projected dispersion of photonic waveguide structure. The grey regions represent the bulk states, while the red solid line delineates the edge state dispersion. (e) The transmission spectra of photonic waveguide state structure. (f) The transmission spectra of waveguide states in a serially connected optical cavity system.

Furthermore, the robust topological interface states serve as an efficient and stable channel that connect multiple biosensors resiliently. We reveal this aspect by studying the disorder effects on the transmission for both the topological and normal biosensors in a configuration with two connected defect-like sensors. With the random disorders similar to that in the single sensor cases in Figs. 4(a) and (e), we show that for the cases with two serially connected sensors, the disorder effect is even stronger for both normal and topological biosensors [Figs. 4(b) and (f)]. For the case of the normal biosensors, the disorder substantially reduces the transmittance away from the resonant frequencies of the two defect-like cavities. The quality factors of the two cavity resonances are also notably reduced as indicated by the broadened dips [Fig. 4(f)]. These features indicate significant reduction of the performance for the serially connected biosensors, which is a key challenge for the integration of biosensors in the technology frontier. In contrast, for the topological biosensors, the effect of the same disorder is almost negligible: The transmission spectrum is only marginally modified, while the quality factors of the two cavity resonances are nearly unchanged. These results encouragingly suggest that the integration of multiple biosensors along the topological interface channel is possible in real working conditions, which is the main finding in this work.

**Photonic crystal biosensor**

Harnessing the attribute of the robust photonic transport in the topological interface channel, we propose a scheme for the topological integration of photonic biosensors, as schematically illustrated in Figs. 1(a)-1(d). The essence here is to use the topological interface channel as a 1D highway for photons to connect multiple photonic biosensors, where each biosensor is based on a defect-like cavity mode with a distinctive resonant frequency. In biomedical sensing applications, each biosensor is designed for a specific biomarker via a unique biorecognition mechanism such as the antibody-antigen binding or the DNA aptamer-protein binding [16,17]. By exploiting many of such topological integration in parallel, one can integrate many biosensors in a single chip which could enable the detection of a large number of biomarkers with a single shot transmission measurement.

We now demonstrate this working principle by studying the transmission spectrum for a system with four defect-like biosensors serially coupled to a single topological interface channel, as depicted in Fig. 1(a). In our calculation, the four defect-like cavity modes that make the biosensors are created by replacing four square rods of the PhC UC1 with four circular rods of the radii $0.126a$, $0.084a$, $0.326a$, and $0.284a$, separately. The choice of these radii enables the creation of four cavities of distinctive frequencies, as shown in the transmission spectrum in Fig. 5 [see also Fig. 1(c)]. For simplicity, we assumed here that these circular rods are of the same dielectric constant as silicon. If otherwise, the radii can be redesigned to achieve the same goal. Figure 1(b) also shows that there is negligible crosstalk between these resonances. Therefore, each biosensor can perform their function independently. The robustness of the topological interface channel ensures the resilient performance of the device against defects and imperfection that are deleterious in real-life applications.

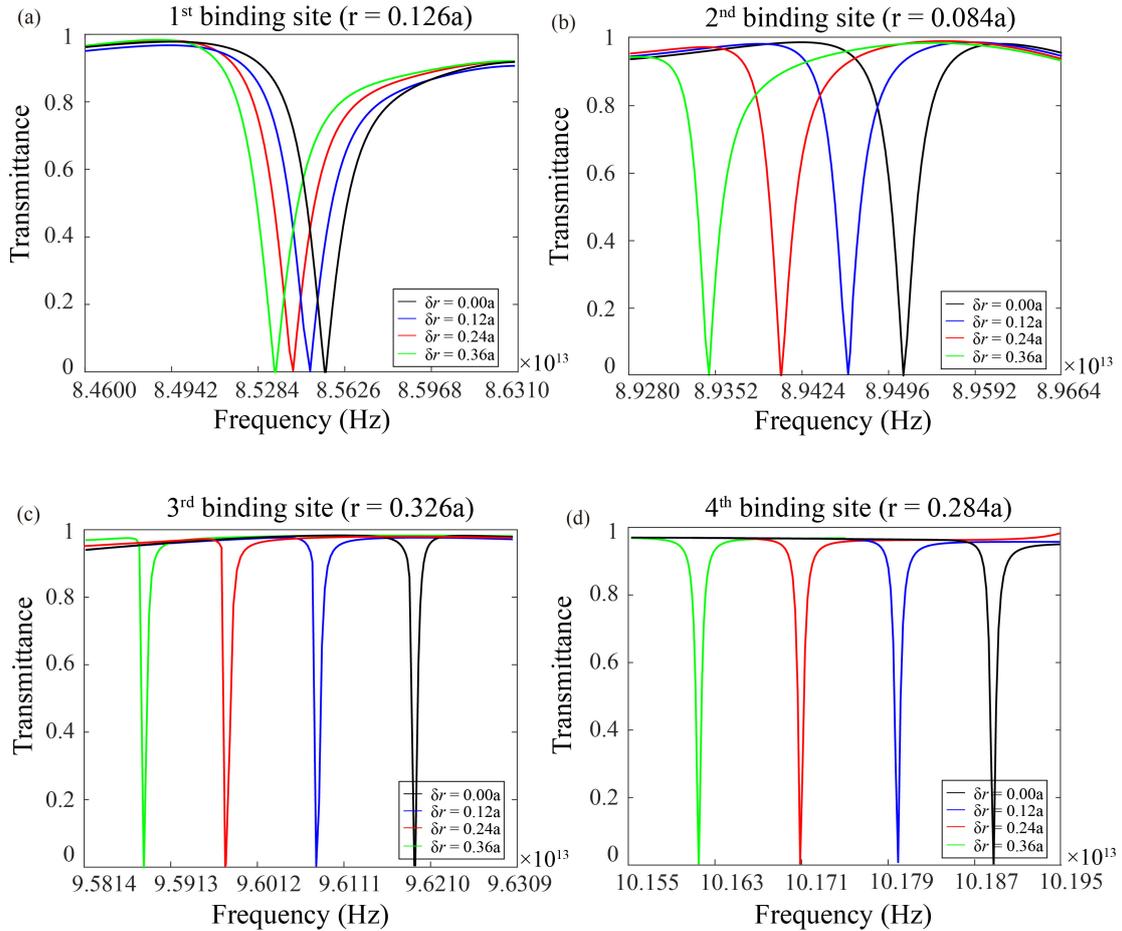

**Fig. 5**. By replacing four square rods with four circular rods which serve as the binding sites for the biomarkers, four biosensors are created and coupled to the topological interface channel serially. The first, second, third, and fourth circular binding sites have the radius $0.126a$, $0.084a$, $0.326a$, and $0.284a$, separately. The global transmission spectrum for the whole photonic band gap is given in Fig. 1(c). Here the local transmission spectrum for each biosensor is shown for various amount of biomarkers bounded to the binding sites which are simulated via the increase of the effective radius of the binding sites, i.e., the change of the effective radius of a binding site $\delta r$ characterizes

the amount of bounded biomarkers. Figures (a)-(d) give the modification of the transmission spectra for the first, second, third, and fourth biosensors, separately, when the effective radius changes as labeled by various $\delta r$.

We now study the quantitative performance of the device via the sensitivity of each biosensor. In general, since the biosensors are designed via a conventional scheme, i.e., via the defect-like cavity mode induced by modifying the rods in the PhC. We believe that the sensitivity of each biosensor is at the same level of previous designs [9, 13, 14, 25]. As shown in Fig. 5, notable shifts of the resonant frequencies can be seen for the four biosensors when biomarkers are bound to the surface of the four circular rods. Here, the biomarker binding is simulated by the increase of the effective radius of each circular rod. The change of the effective radius is denoted by $\delta r$. As shown in Fig. 5, the biomarker binding at each biosensor, as indicated by various $\delta r$, can be revealed by the shift of the resonant frequency (i.e., the dip) in the transmission spectrum. Such shift in turn gives a pathway to quantify the amount of bounded biomarkers through the transmission spectrum. Since each biosensor has a distinctive resonant frequency and binds a particular type of biomarker, by analyzing the shift of the resonant frequencies, one can quantitatively determine the amounts of four different types of biomarkers with a single shot transmission measurement.

For such quantitative analysis of the bounded biomarkers, we further analyze the dependences of the shift of the resonant frequencies on the change of the effective radii of the four binding sites, $\delta r_i$ for $i = 1,2,3,4$. The calculated results are presented in Fig. 6. By using the linear regression of the data, we obtain the following quantitative dependence for the four biosensors: $f_1 = (-0.05\delta r_1 + 8.556) \times 10^{13}$ Hz, $f_2 = (-0.05\delta r_2 + 8.952) \times 10^{13}$ Hz, $f_3 = (-0.0833\delta r_3 + 9.618) \times 10^{13}$ Hz, and $f_4 = (-0.0833\delta r_4 + 10.188) \times 10^{13}$ Hz. The slopes of the lines in Fig. 6 reflect the sensitivity of the biosensors. Note that the true sensitivity should also consider the radii of the four binding sites are different. That is, the sensitivity should be quantified based on the change of the area of a binding site instead of the radius. Besides, in real-life applications, the sensitivity is also limited by the quality factor of the cavity mode for each biosensor which can be reduced by various photon loss mechanisms. Nevertheless, the sensitivity of these biosensors is comparable with previous studies [9, 25], while the integration scheme here is more efficient and robust.

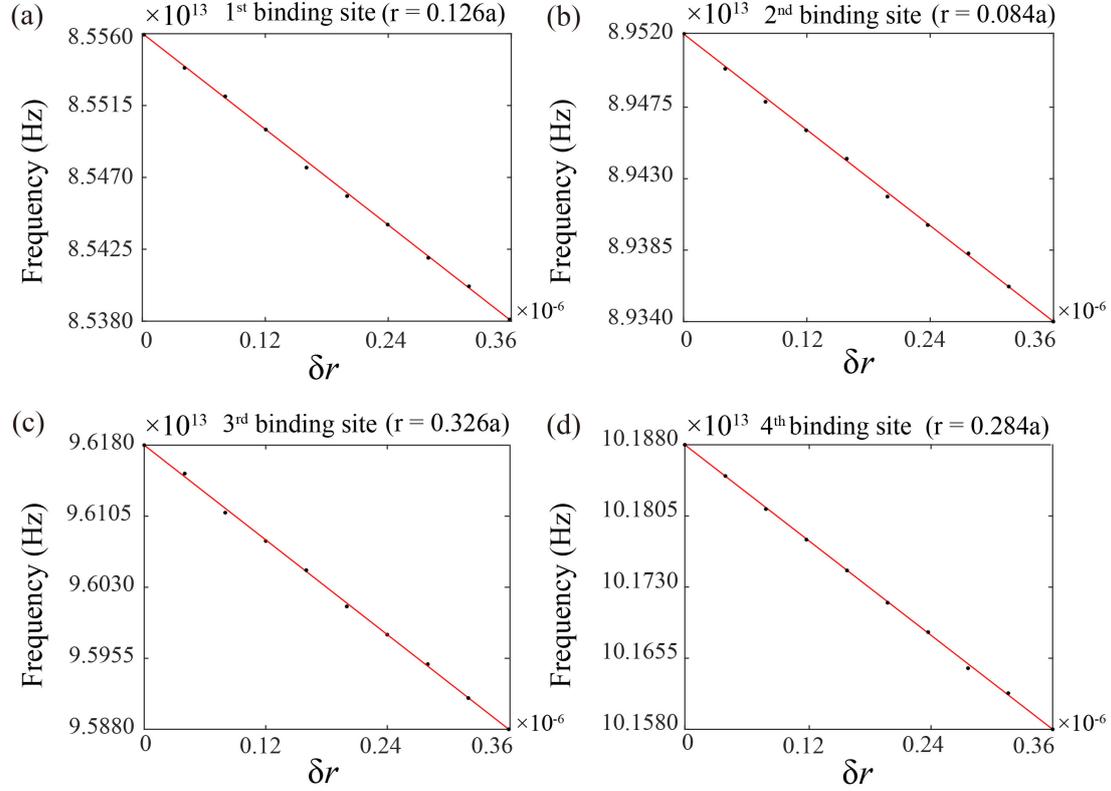

**Fig. 6.** (a)-(d) The resonant frequency as a function of the analyte binding thickness, i.e., the change of the effective radius of each binding site, at the four biosensors, respectively. The solid circles represent electromagnetic simulation results, while the lines represent the fitting results.

**Conclusion and outlook**

Inspired by the robustness of the topological edge transport of photons, we propose here a topological scheme for the integration of multiple photonic biosensing functions where many photonic biosensors based on defect-like cavity modes in PhCs are coupled to the topological interface channels between two types of PhCs. We first reveal that such a cavity-interface coupling gives rise to robust transmission spectrum and hence resilient performance in photonic biosensing against disorders. In contrast, we find that for photonic biosensors coupled with the conventional PCW channel, the same disorder effect leads to substantial modification of the transmission and is thus deleterious for biosensing applications in reality. Furthermore, we uncover that when the topological interface channel is coupled with multiple photonic biosensors, the cross talk between these biosensors can be negligible and the performance of the entire device is robust again disorders as well. This scheme with multiple photonic biosensors offers a way to integrate many photonic biosensors in a single photonic crystal chip and thus realizing multiplexing biosensing. By exploiting both the serial and parallel couplings to many topological interface channels, our scheme can provide a pathway for the integration of many photonic biosensors. Therefore, the scheme proposed here provides a promising route toward reliable large-scale integration of on-chip biosensors with reliable sensing functions. Considering the simpleness and effectiveness of our design, we believe that the proposed topological integration of photonic biosensors could be valuable for future high-throughput biosensing technology. The proposed scheme could also be useful for the realization of integrated multiplexing sensing of molecules or chemicals with high efficiency based on various binding mechanisms.

**Acknowledgement**

This work was supported by the National Natural Science Foundation of China (Grant Nos. 12125504 and 12074281), the "Hundred Talents Program" of the Chinese Academy of Sciences, and the Priority Academic Program Development (PAPD) of Jiangsu Higher Education Institutions.